\documentclass[a4paper,UKenglish,cleveref,autoref]{oasics-v2021}
\usepackage{tikz}

\title{Extending the Time Horizon: Efficient Public Transit Routing on Arbitrary-Length Timetables}
\titlerunning{Efficient Public Transit Routing on Arbitrary-Length Timetables}

\author{Sascha Witt}{Karlsruhe Institute of Technology, Germany}{sascha.witt@kit.edu}{https://orcid.org/0000-0002-7867-3200}{}
\authorrunning{S.\,Witt}
\Copyright{Sascha Witt}


\nolinenumbers
\hideOASIcs

\newcommand*{\mleft}{\mathopen{}\mathclose\bgroup\left}
\newcommand*{\mright}{\aftergroup\egroup\right}

\newcommand*{\chTime}[1]{\Delta\tau_{\mathrm{ch}}\mleft(#1\mright)}
\newcommand*{\fpTime}[2]{\Delta\tau_{\mathrm{fp}}\mleft(#1,#2\mright)}
\newcommand*{\nthStopOn}[2]{#1\textrm{\scriptsize @}#2}
\newcommand*{\arrTime}[2]{\tau_{\mathrm{arr}}\mleft(\nthStopOn{#1}{#2}\mright)}
\newcommand*{\depTime}[2]{\tau_{\mathrm{dep}}\mleft(\nthStopOn{#1}{#2}\mright)}
\newcommand*{\transfer}[4]{\nthStopOn{#1}{#2} \rightarrow \nthStopOn{#3}{#4}}

\begin{document}

\maketitle

\begin{abstract}
We study the problem of computing all Pareto-optimal journeys in a public transit network regarding the two criteria of arrival time and number of transfers taken.

In recent years, great advances have been made in making public transit network routing more scalable to larger networks.
However, most approaches are silent on scalability in another dimension: Time.
Experimental evaluations are often done on slices of timetables spanning a couple of days, when in reality, the planning horizon is much longer.

We introduce an extension to trip-based public transit routing, proposed in~\cite{Witt2015}, that allows efficient handling of arbitrarily long timetables.
Our experimental evaluation shows that the resulting algorithm achieves fast queries on year-spanning timetables, and can incorporate updates such as delays or changed routes quickly even on large networks.
\end{abstract}

\section{Introduction}
\label{sec:intro}
Although superficially similar, finding journeys in public transit networks is very different from finding routes in road networks.
This is evident from the difficulties in adapting algorithms for road networks to public transit networks~\cite{Berger2009}.
Two of the main reasons for this are time-dependency and multiple criteria.

Arguable the most common criteria when planning routes in both road and public transit networks is travel time.
For road networks, this is often sufficient, but not so for public transit networks.
While multiple criteria are not a requirement for public transit routing algorithms, there is one criteria besides travel time that many travellers care about:
  The number of required transfers.
This is because changing between trains is inconvenient, especially when travelling with luggage.
Additionally, there's the risk of missing a connection when the earlier train is delayed.

Therefore, most algorithms for public transit routing are multi-criteria in at least travel time and the number of transfers, and return a set of Pareto-optimal results.
The user can then choose between faster or more convenient travel.
Although multi-criteria Pareto optimization is NP-hard in general, this particular combination of criteria is efficiently tractable in public transit networks~\cite{Muller-Hannemann2006}.

The more important aspect of public transit networks is the inherent time-dependency.
While road networks can be time-dependent when one considers traffic data or road closures, it is one of the core properties of public transit networks.
Travel is only possible at certain times; in more remote areas, travel may even be impossible during the night or on certain days.

This time-dependency has an important implication:
An algorithm can only answer queries that fall within the time period for which it has data.
Many state-of-the-art algorithms require preprocessing to enable fast queries.
If this preprocessing is too costly, it may not be feasible to process a long time period, which means users may not be able to plan journeys in advance.

\subsection{Related Work}
\label{sec:related}
While many public transit routing algorithms have been proposed, especially in recent years, to the best of our knowledge, none of them explicitly tackle the problem of long (in the time sense) timetables.

For an extensive survey of both classic and modern routing algorithms, please refer to Bast et~al.~\cite{Bast2014a}.

Pyrga et~al.~\cite{Pyrga2008} solve the problem of finding optimal journeys by finding shortest paths in graphs.
They introduce the time-extended and time-dependent model as well as speed-up techniques such as goal directed search.
In general, however, running variants of Dijkstra's algorithm on these graph structures is outperformed by current state-of-the-art techniques.

Berger et~al.~\cite{Berger2010} propose SUBITO and k-flags, two speed-up techniques that Pareto-optimize both travel time and number of transfers.
Their techniques allow multi-criteria queries with dynamic update, but have the restriction that they cannot add a new arc to the network.
As they note, this is not critical in train networks, but if one also considers buses (especially as rail replacements), new arcs are much more likely to appear in updates.
Their timetable seems to span a single day.

RAPTOR~\cite{Delling2012} does not explicitly model the data as a graph, instead operating directly on the timetable data.
Their experiments cover the usual single day, although their model suggests that it may be able to be extended to longer periods.

The Connection Scan Algorithm (CSA)~\cite{Dibbelt2018} also foregoes graphs and instead works on an ordered array of connections.
This simple model makes it easy to adjust to changes in the timetable, but it is not obvious if this still holds for large timetables.
In addition, CSA's performance relies on efficient use of CPU caches, and cache efficiency is expected to degrade with increasing amounts of data.

Public Transit Labelling (PTL)~\cite{Delling2015} uses a hub labelling approach.
An existing hub labelling algorithm is used to compute departure and arrival labels for each stop, which can, at query time, be used to reconstruct optimal journeys.
Those labels are then optimized by exploiting the properties of public transit networks and queries.
PTL achieves very low query times, even for multi-criteria queries, at the cost of extensive preprocessing and high memory footprint.

Transfer Patterns (TP)~\cite{Bast2010,Bast2016,Bast2014} is a speed-up technique that precomputes transfer patterns between all stops in the network.
Transfer patterns are defined as the sequence of stops where passengers transfer between vehicles.
At query time, these patterns are applied to efficiently find all Pareto-optimal journeys.
Similar to this work, TP uses bit sets to indicate traffic days for trips, although it is not clear from the original publication how many days are considered.
[Bast2014] specifies that only single day and seven day periods are evaluated.

\subsection{Our Contribution}
\label{sec:contribution}
In this work, we present how the trip-based public transit routing algorithm first introduced in~\cite{Witt2015} can be modified to be able to efficiently handle time periods of arbitrary length.
Section~\ref{sec:prelim} introduces necessary definitions, before Section~\ref{sec:basic} gives a high-level introduction to the basic trip-based algorithm.
Section~\ref{sec:extend} proposes changes to the algorithm for efficient handling of large time periods,
  with Section~\ref{sec:updating} describing how the data can be updated without complete re-computation when changes to the timetable occur.
Finally, Section~\ref{sec:experiment} contains the experimental evaluation of the proposed changes.

\section{Preliminaries}
\label{sec:prelim}
This section introduces required definitions and notation used in the remainder of this work.

We define a public transit network via an aperiodic \textbf{timetable}, which consists of a set of stops, a set of footpaths, and a set of trips.

A \textbf{stop} is a location where passengers can enter or exit a vehicle, such as a bus or train.
A stop may represent anything between an entire train station and a single platform, or even part of a platform, depending on the granularity of the model.
Less fine-grained models often assign a \textbf{minimum change time} $\chTime{\cdot}$ to each stop.
This is the minimum time that must pass between the arrival of one vehicle and the departure of another for passengers to be able to transfer between them.
This tries to account for the time people need to go from their arrival platform to the departure platform.
It can also serve as buffer time to prevent minor delays from making a journey unfeasible.

A \textbf{footpath} connects two stops.
A footpath indicates that passengers are expected to travel between these two stops on foot.
This may be between two nearby stations, or -- on more fine-grained models -- between two platforms at the same station.
We use the most general model of directed, non-transitive footpaths, which means that walking is only considered between stops directly linked via a footpath.
The time required to walk from stop $s_1$ to stop $s_2$ is denoted by $\fpTime{s_1}{s_2}$.
If no footpath between $s_1$ and $s_2$ exists, we define $\fpTime{s_1}{s_2} := \infty$, for ease of definitions.

A \textbf{trip} represents a vehicle, usually a bus, train, or similar.
Each trip travels at a specific time along a sequence of stops, denoted by $\left<\nthStopOn{t}{1},\nthStopOn{t}{2},\ldots,\nthStopOn{t}{n}\right>$.
For each $\nthStopOn{t}{i}$, the timetable contains the arrival time $\arrTime{t}{i}$ and the departure time $\depTime{t}{i}$ of trip $t$ at that stop.
We define $\arrTime{t}{1} := 0$ and $\depTime{t}{n} := \infty$.
Note that a trip may visit the same stop more than once, such as on circular routes.

Trips are partitioned into \textbf{routes}, such that all trips belonging to the same route visit the same sequence of stops and none overtakes any other
 -- no trip departs later at one stop and arrives earlier at a later stop than any other trip of the same route.
More specifically, we require that all trips of a route can be totally ordered with respect to
\begin{equation}
  t_1 \preceq t_2 \iff \forall i \in \mleft[1, n\mright]:
    \arrTime{t_1}{i} \leq \arrTime{t_2}{i} \land
    \depTime{t_1}{i} \leq \depTime{t_2}{i} \text{.}
\end{equation}
Trips that violate this requirement are assigned to different routes, even though they share the same stop sequence.
This ``no overtaking'' rule ensures that it is never beneficial to wait for a later trip of the same route.
Note that these routes do not necessarily correspond to any e.g.\ subway lines as defined by the transport provider, they are simply a way to group trips in a way useful for route planning.

A \textbf{transfer} is a connection between two trips.
Whenever passengers can exit one trip and board another, subject to the constraints given by footpaths and minimum change times, we say that a transfer exists.
This means that for each transfer $\transfer{t_1}{e}{t_2}{b}$,
\begin{equation}
  \arrTime{t_1}{e} + \chTime{\nthStopOn{t_1}{e}} \leq \depTime{t_2}{b}\label{eqn:transfer_at_stop}
\end{equation}
must hold; if the transfer involves a footpath, the requirement is instead
\begin{equation}
  \arrTime{t_1}{e} + \fpTime{\nthStopOn{t_1}{e}}{\nthStopOn{t_2}{b}} \leq \depTime{t_2}{b} \text{.}\label{eqn:transfer_via_footpath}
\end{equation}

A \textbf{journey} is, essentially, a sequence of trips and interspersed footpaths.
The result of a query is one or more journeys telling the user how to get from their source stop to their destination.

\section{Basic Trip-Based Public Transit Routing}
\label{sec:basic}
This section describes the basic trip-based public transit routing algorithm first introduced in \cite{Witt2015}.

The core idea of the algorithm is to put the focus on the fundamental building blocks of public transit networks, the trips.
This is most apparent during the query phase:
Whereas other algorithms track progress by recording which stops can be reached, the trip-based algorithms records which \emph{trips} can be reached.

\subsection{Preprocessing}
Before queries can be performed, a short preprocessing phase is required.
The preprocessing can itself be divided into two steps: Transfer computation and transfer reduction.

During transfer computation, possible transfers between pairs of trips are identified.
The process is fairly straightforward: For each trip $t$, we iterate over all its stops.
At each stop $\nthStopOn{t}{i}$ we find, for each route $r$ visiting that stop, the first trip $u$ that can be reached from our current trip, i.e., the earliest trip such that equation \ref{eqn:transfer_at_stop} is satisfied.
If there is such a trip, we record the existence of a transfer $\transfer{t}{i}{u}{j}$, where $j$ is the index of that stop in the stop sequence of $u$.
We then repeat this for each stop reachable via footpath from $\nthStopOn{t}{i}$, except that equation \ref{eqn:transfer_via_footpath} must be satisfied.
Since each trip is processed independently, this is trivially parallelizable.

There are two main advantages to this reification of transfers.
Firstly, since all possible transfers are identified during preprocessing, the query algorithm does not need to evaluate minimum change times or examine footpaths.
Secondly, this allows arbitrarily complex transfer constraints.
For example, it may be known that two routes always stop at neighbouring platforms, so the minimum change time at this stop may be lower for transfers between trips of these specific routes.
Any constraints (or relaxations) that affect the existence of a transfer is dealt with during this preprocessing step, and has therefore no further performance cost at query time.

The second, optional preprocessing step is transfer reduction.
In this step, transfers which do not lead to non-dominated journeys are removed from the previously computed set.
This significantly reduces the number of transfers, which reduces memory usage and improves query performance.

As previously, we process each trip independently, allowing trivial parallelization.
For each trip, we keep track of reached stops together with arrival time, as well as reached trips together with the index in their stop sequence where we reach them.
Stops are processed in reverse order, starting with the last stop visited by the trip.
At each stop $\nthStopOn{t}{e}$, we evaluate each transfer $\transfer{t}{e}{\cdot}{\cdot}$.
Transfers which improve arrival times or allow us to reach an earlier trip, or an earlier stop on a trip, are kept.
All other transfers are removed.

\subsection{Query}
The query algorithm is reminiscent of a breadth-first search, with trips as graph nodes and transfers as edges between them.

First, we identify routes that visit the destination stop or stops connected to the destination stop via a footpath.
These are the destination routes and are recorded, together with the stop index where passengers should exit and the length of the footpath, if applicable.

For an earliest-arrival query, we then find, for each route at the source stop or a stop reachable via footpath from the source stop, the first reachable trip.
These trips, and the stop indices where they are reached, are put into a queue.
The entries $(t, i)$ of this queue are then processed as follows.

First, if we already found a journey, we might be able to prune this entry.
To do so, we compare the previously found arrival time at the destination stop with $\arrTime{t}{i+1}$.
If this entry's time is later, it clearly cannot lead to an improved journey, and can therefore be skipped.
Otherwise, we check if this entry's trip belongs to one of the destination routes and reaches the destination.
If so, we update the arrival time at the destination stop using the recorded data for this route.

Next, we examine all transfers $\transfer{t}{j}{\cdot}{\cdot}$, with $j > i$.
If a transfer leads to an earlier trip of a route or an earlier stop on a trip than previously reached, a new entry is added to the queue.
To determine this, we keep -- for each route -- track of the entries that were previously added to the queue.

Once the queue is empty, we report the non-dominated journeys found.

A profile or range query uses the procedure described above as a subroutine.
We start by finding all reachable trips in the given departure time interval and ordering them by departure time.
We then process them in reverse order, starting with the latest possible departure.
For each distinct departure time, we perform an earliest-arrival query as described above.
However, we do not reset the tracking data structure between each of these.
This means that each (earlier) departure will not re-explore the paths taken by later departures, which would only lead to dominated journeys.
For this to work, we require only small changes to the algorithm described above:
We now also track the number of transfers required to reach a trip, and also compare against that when determining whether a transfer leads to a new entry.
In addition, the pruning rule now also takes the number of transfers into account, because a journey with a later arrival time may still be optimal if it requires fewer transfers.

\section{Extending the Time Horizon}
\label{sec:extend}
Although in theory, the basic trip-based algorithm can be applied to timetables of arbitrary time intervals, it is not efficient to do so.
The reason for this is that while timetables are aperiodic, they usually do have some regularity:
Many trips operate at the same time on most days, with exceptions for weekends, holidays, etc.
Treating each instance of this trip as independent is correct, but not optimal.
This section introduces enhancements to the trip-based algorithm to allow it to efficiently handle large time intervals, such as timetables spanning an entire year.

\subsection{Changes to the Model}
\label{sec:extend_model}
The underlying model of the timetable remains largely unchanged, except for one crucial addition:
Each trip now also has an associated bit set that indicates on which days the trip is active,
  i.e., a $1$ at the $i$-th position of the set means that the trip operates on the $i$-th day of whatever time interval the timetable covers.

\subsection{Preprocessing}
\label{sec:extend_preprocessing}
Preprocessing is extended to account for the new bit sets.
For transfer computation, this simply means that transfers are only allowed when both trips have at least one active day in common.
One consequence of this is that a trip may have multiple transfers to different trips of the same route, because the earliest reachable trip may not be active on all days.

For transfer reduction, the changes are slightly more complicated.
Here, we additionally keep track of the days on which each transfer is required, again by using a bit set.
During the query stage, we can then select only those transfers that are valid during the requested day(s).
This increases memory usage, but improves query performance.

\subsection{Query}
\label{sec:extend_query}
Although the query stage works the same conceptually, there are some differences in how the data is managed.

The query now must also specify a date on which the journey should take place.
The first step then is to extract the reduced transfers for this date and the following days, in case of long journeys.
This adds some runtime overhead, however, the extracted data can easily be cached, which avoids paying this cost repeatedly.

After this, we proceed to find reachable trips as before, paying attention to the days of operation of each trip.
The remainder of the query stage stays the same -- we don't need to pay any more attention to the bit sets, because we only look at valid transfers, which means that only active trips are reached.

The main challenges are on the engineering side, where data and code has to be structured differently to account for the altered model.
The most important change is in how trips are identified: Whereas before, trips were simply numbered from $0$ to the total number of trips, it's now more complicated, since a query might encounter more than one instance of a trip.
It's also unfeasible when updating the timetable data, which we will discuss in the next section (\ref{sec:updating}).

Instead of a simple integer, trips are now identified by a triple of route ID, trip ID, and a day offset.
The trip IDs are scoped to each route and ordered according to $\cdot\leq\cdot$ (i.e., time).
The day offset is relative to the query day: The day given by the query corresponds to $1$, the next day to $2$, etc.
An offset of $0$ is used for the day before the query, to allow the use of trips that start on the previous day, but cross into the current one.
The trip ID and the day offset is stored as a single integer, with the offset in the higher bits.
This enables efficient comparisons when testing which trip is earlier.

However, since the number of trips is no longer fixed, some of the optimizations used in~\cite{Witt2015} no longer work.
In particular, the original paper proposed ``unrolling'' the data structure used to track reached trips, which makes updates slightly slower (because all following trips also have to be updated),
  but lookups much faster (because only a single, known memory location has to be read).
While we could apply the same trick in this work if we limited the query to a low enough number of days, we chose not to.
Instead, we use explicit (Pareto-)sets using tuples of trip ID, stop index, and number of transfers.
This allows us to be more flexible, both in the number of days a query may span, and with regard to possible extensions with more optimization criteria.

Overall, the more complex data layout comes with a small performance cost, compared to the more streamlined original version.
However, it is easy to extract arbitrary time periods into the flat data structures used by the original algorithm, trading extra memory for reduced running time.
If desired, one could adopt a hybrid model in practice, were frequently queried periods (such as the next few days) use the flat structures, while the rest is served by the algorithm proposed in this work.

\section{Updating Timetable Data}
\label{sec:updating}
With an increased time period covered by the timetable, it is likely that occasional changes are required.
Construction work or outages may close some stops and add new ones, require alternate routes for trip, or disable some routes completely.
On top of that, delays and outages may happen no matter how long the time period is.
For short timetables, a complete reprocessing of the data may be feasible, but for longer timetables, we can do better.

While there are many different types of alterations possible, the two fundamental ones are adding and removing trips.
For example, a delay can be incorporated by removing a trip and re-adding it with updated arrival and departure times.

Removing a trip requires updating trips that had a transfer to that trip.
Adding a trip requires updating trips that may have a transfer to the new trip, in addition to computing transfers from the new trip.
In our experiments, we simply re-compute and re-reduce all transfers from the updated trips.
More fine-grained updates may be possible, but would require more sophisticated algorithms to identify which transfers need to be added or removed.

\section{Experimental Evaluation}
\label{sec:experiment}
We ran experiments to measure the required time for preprocessing and queries, as well as for dynamic updates.
We performed our experiments using a 64-core AMD Epyc 7702P processor and 1024\,GB DDR4 RAM.
Preprocessing and each individual dynamic update is done in parallel, using up to 128 threads.
Queries are performed sequentially.

We evaluated five different real-world data sets, covering public transit networks of varying sizes:
Germany, provided to us by Deutsche Bahn, Switzerland, (acquired from \texttt{gtfs.cheops.ch}), the Netherlands (\texttt{gtfs.ovapi.nl}), Sweden (\texttt{trafiklab.se}), and Madrid (\texttt{emtmadrid.es}).
These data instances are summarised in Table~\ref{tab:data_sets}.

\begin{table}[tp]
  \centering
  \caption{Instances used for experiments.}\label{tab:data_sets}
  \begin{tabular}{l r r r r r r}
    \hline
    Instance    & Days & Stops & Trips     & Routes & Footpaths \\\hline
    Germany     & $363$ & $249$\,k & $1\,703$\,k & $240$\,k & $400$\,k \\
    Switzerland & $363$ &  $30$\,k &    $959$\,k &  $52$\,k &  $94$\,k \\
    Netherlands & $145$ &  $73$\,k &    $636$\,k &  $17$\,k & $329$\,k \\
    Sweden      &  $97$ &  $51$\,k &    $257$\,k &  $22$\,k &  $61$\,k \\
    Madrid      & $372$ & $4.6$\,k &    $193$\,k & $1.4$\,k &  $22$\,k \\
    \hline
  \end{tabular}
\end{table}

Figures for preprocessing can be found in Table~\ref{tab:preprocessing}.
We can see that the time required for transfer computation is negligible.
Transfer reduction takes a noticeable, but still fairly low amount of time, and results in a substantially lower number of transfers,
  which affects both memory usage and query performance positively.

\begin{table}[tp]
  \centering
  \caption{Preprocessing time and number of transfers.
    Listed are the total number of transfers computed, the number of transfers remaining after reduction,
    the ratio of the two, and the time required to compute and reduce the transfers (in parallel).
  }\label{tab:preprocessing}
  \begin{tabular}{l r r r r r r}
    \hline
    Instance & Total transfers & Reduced transfers & Ratio & Comp.\ time & Reduction time \\\hline
    Germany     & $2\,357$\,million & $335$\,million & $14.2\,\%$ & $15$\,s & $26.2$\,min \\
    Switzerland &    $903$\,million & $167$\,million & $17.4\,\%$ & $11$\,s & $14.1$\,min \\
    Netherlands &    $544$\,million &  $78$\,million & $14.4\,\%$ &  $5$\,s &  $3.6$\,min \\
    Sweden      &    $247$\,million &  $51$\,million & $21.0\,\%$ &  $2$\,s &  $1.3$\,min \\
    Madrid      &    $442$\,million &  $37$\,million &  $8.5\,\%$ &  $8$\,s &  $1.3$\,min \\
    \hline
  \end{tabular}
\end{table}

Before we can perform queries, we need to extract the transfers relevant to that query, as described in Section~\ref{sec:extend_query}.
The evaluations of this process can be found in Table~\ref{tab:preparation}.
When executed in parallel, the total time for extraction is fairly negligible -- at most up to $10$ seconds for Germany.
Ideally, this extraction is done once in advance and then cached, which is what we did for our query experiments.
Of course, this requires a certain amount of memory, but not unreasonably so.
If RAM is short, one could use, e.g., an LRU caching strategy.

The final column in Table~\ref{tab:preparation} lists the time required to transform the data so it can be consumed by the original trip-based algorithm proposed in~\cite{Witt2015}.
This ``flat'' format is simpler and allows a more efficient query, as we will see next.

\begin{table}[tp]
  \centering
  \caption{Query preparation.
    Listed are the (sequential) times required to extract transfers for all days and for a single day,
    followed by the total memory usage for all extracted transfers.
    The final column lists the time required to ``flatten'' two consecutive days so the original trip-based algorithm~(\cite{Witt2015}) can be used.
  }\label{tab:preparation}
  \begin{tabular}{l r r r r r r}
    \hline
    Instance & Total extr.\ time & Extr.\ time per day & Memory usage & Flatten time \\\hline
    Germany     & $6.7$\,min & $1\,110$\,ms & $163$\,GB & $3\,096$\,ms \\
    Switzerland & $2.3$\,min &    $377$\,ms &  $37$\,GB &    $645$\,ms \\
    Netherlands &    $30$\,s &    $207$\,ms & $9.6$\,GB &    $400$\,ms \\
    Sweden      &    $18$\,s &    $186$\,ms & $7.4$\,GB &    $505$\,ms \\
    Madrid      &    $56$\,s &    $151$\,ms &  $52$\,GB &    $599$\,ms \\
    \hline
  \end{tabular}
\end{table}

For query evaluation, we performed $10\,000$ full-day profile (24 hours from midnight to midnight) queries, with source stop, destination stop, and query day chosen uniformly at random.
However, not all of these queries can be answered with a valid journey, because at any given time, the network may not connected.
This is because some stops may be seasonable, e.g., only operated in summer or in winter, or they may be relocated stops that only exist during a time of construction at the actual stop location.
These kinds of queries can often be answered very quickly with a negative (empty) result.
In order to not skew the results too much, we disregard queries that do not return at least one valid journey in the following.

\begin{figure}[tbp]
  \centering
  \input{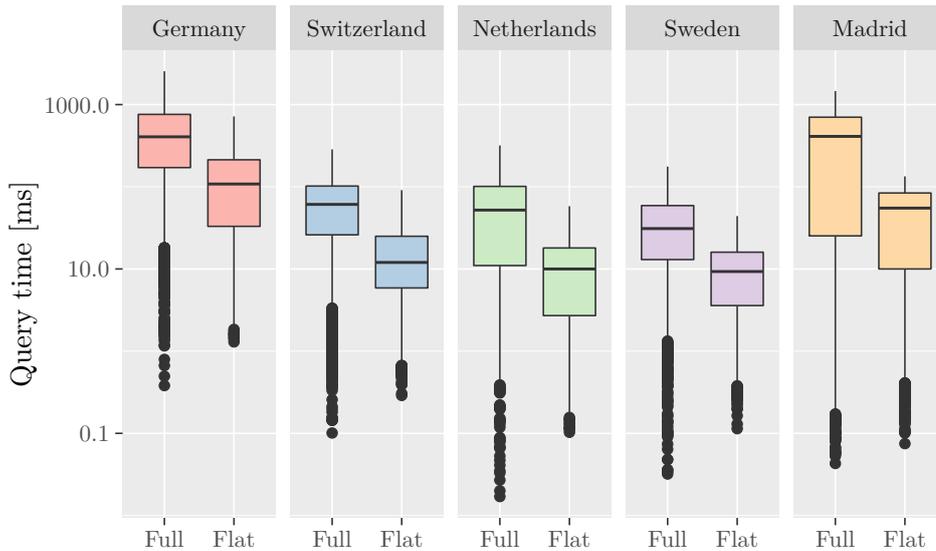}
  \caption{Query times, on logarithmic scale.
    \texttt{Full} refers to queries on the full timetable as presented in this paper.
    \texttt{Flat} refers to queries using the original trip-based algorithm on a relevant portion of the timetable (cf.~\ref{sec:extend_query}).
  }\label{fig:query_eval}
\end{figure}

The results of the query evaluation can be found in Figure~\ref{fig:query_eval}.
Note that the y-axis scale is logarithmic.
The left box plot for each instance corresponds to the queries performed on the full timetable as described in this paper.
For comparison, the right box plot shows the query times using the original algorithm.
The factor between the two is around $4$, with Madrid a notable outlier at $8$.
The reason for the difference in execution time is that the original algorithm has much simpler (flatter) data structures, which allows more streamlining in the query execution.
The more sophisticated data structures used in this work are used to keep the data more compact, and to make updates to that data more efficient.
It's worth noting that queries on the full timetable finds some journeys that the flat version misses, because we limited each query to two consecutive days.

For dynamic updates, we delay trips, chosen uniformly at random, by removing them, adding a random delay, and adding them again.
While this is only one kind of updates amongst many possible ones, as noted in Section~\ref{sec:updating}, most changes to the timetable can be implemented by removing and adding trips.
Furthermore, the costly part of doing updates is not the actual change itself, but the re-computation and re-reduction of transfers of neighbouring trips, which may gain or lose transfers to new or removed trips.

Figure~\ref{fig:update_eval} shows the time required for such an update for each instance.
Madrid is the most expensive one, and the only one where a majority of updates took multiple seconds.
This is because unlike the other four instances, Madrid is a metropolitan network, and much denser than the others.
Therefore, each trips has many neighbours, so each update touches more trips.
This becomes clearer when we normalize the values by the number of trips for which we recompute the transfers~(Figure~\ref{fig:update_eval2}).

A consequence of this is that batched updates of a local part of the network are much more efficient than individual ones.
When making multiple changes to neighbouring trips, it is likely that the sets of trips requiring new transfers have large overlap.
By delaying the computation of these transfers until all pending changes have been applied, updates can become much more efficient.

\begin{figure}[tbp]
  \centering
  \begin{minipage}[t]{0.48\textwidth}%
    \input{update_times.fig}%
    \caption{Update times, on logarithmic scale.}\label{fig:update_eval}%
  \end{minipage}\hfill%
  \begin{minipage}[t]{0.48\textwidth}%
    \input{update_time_per_trip.fig}%
    \caption{Update times, normalized to show time per updated trip, on logarithmic scale.}\label{fig:update_eval2}%
  \end{minipage}
\end{figure}

\section{Conclusion}
\label{sec:conclusion}
We introduced an extension of the trip-based public transit routing algorithm presented by~\cite{Witt2015}, to allow efficient covering of timetables of arbitrary length.
Our experiments showed that this increased flexibility comes at a runtime cost.
However, this gap can be closed by falling back on the original algorithm for hotspots where this is beneficial, by quickly transforming the necessary data.

An important aspect when covering larger periods of time is the ability to update the data without starting from scratch.
We showed that thanks to careful modelling of the data, a simple approach achieves update performance acceptable for real-time scenarios.

An interesting open problem is to combine this work with the speed-up technique proposed by~\cite{Witt2016}.

\bibliography{references}

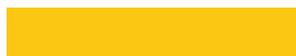
\end{document}